\documentclass[default]{aastex7}
\usepackage{longtable}
\usepackage{multirow}
\usepackage{caption}
\usepackage{booktabs}
\usepackage{amsmath}
\usepackage[symbol]{footmisc}

\begin{document}

\title{In-flight performance of the IXPE telescopes}

\author[0000-0003-1074-8605]{Riccardo Ferrazzoli}
\affiliation{INAF Istituto di Astrofisica e Planetologia Spaziali, Via del Fosso del Cavaliere 100, 00133 Roma, Italy}
\email[show]{riccardo.ferrazzoli@inaf.it}  

\author[0000-0003-4925-8523]{Enrico Costa}
\affiliation{INAF Istituto di Astrofisica e Planetologia Spaziali, Via del Fosso del Cavaliere 100, 00133 Roma, Italy}
\email{enrico.costa@inaf.it}

\author[0000-0003-1533-0283]{Sergio Fabiani}
\affiliation{INAF Istituto di Astrofisica e Planetologia Spaziali, Via del Fosso del Cavaliere 100, 00133 Roma, Italy}
\email{sergio.fabiani@inaf.it}

\author[0000-0002-3638-0637]{Philip Kaaret}
\affiliation{NASA Marshall Space Flight Center, Huntsville, AL 35812, USA}
\email{philip.kaaret@nasa.gov}

\author[0000-0002-1868-8056]{Stephen L. O'Dell}
\affiliation{NASA Marshall Space Flight Center, Huntsville, AL 35812, USA}
\email{steve.o'dell@nasa.gov}

\author[0000-0003-1548-1524]{Brian D. Ramsey}
\affiliation{NASA Marshall Space Flight Center, Huntsville, AL 35812, USA}
\email{brian.ramsey@nasa.gov}

\author[0000-0002-7781-4104]{Paolo Soffitta}
\affiliation{INAF Istituto di Astrofisica e Planetologia Spaziali, Via del Fosso del Cavaliere 100, 00133 Roma, Italy}
\email{paolo.soffitta@inaf.it}

\author[0000-0002-9785-7726]{Luca Baldini}
\affiliation{Istituto Nazionale di Fisica Nucleare, Sezione di Pisa, Largo B. Pontecorvo 3, 56127 Pisa, Italy}
\email{luca.baldini@pi.infn.it}

\author[0000-0002-2469-7063]{Ronaldo Bellazzini}
\affiliation{Istituto Nazionale di Fisica Nucleare, Sezione di Pisa, Largo B. Pontecorvo 3, 56127 Pisa, Italy}
\email{ronaldo.bellazzini@pi.infn.it}

\author[0000-0003-0331-3259]{Alessandro Di Marco}
\affiliation{INAF Istituto di Astrofisica e Planetologia Spaziali, Via del Fosso del Cavaliere 100, 00133 Roma, Italy}
\email{alessandro.dimarco@inaf.it}

\author[0000-0001-8916-4156]{Fabio La Monaca}
\affiliation{INAF Istituto di Astrofisica e Planetologia Spaziali, Via del Fosso del Cavaliere 100, 00133 Roma, Italy}
\affiliation{Dipartimento di Fisica, Universit´a degli Studi di Roma Tor Vergata, Via della Ricerca Scientifica 1, 00133 Rome, Italy}
\email{fabio.lamonaca@inaf.it }

\author[0000-0002-0984-1856]{Luca Latronico}
\affiliation{Istituto Nazionale di Fisica Nucleare, Sezione di Torino, Via Pietro Giuria 1, 10125 Torino, Italy}
\email{luca.latronico@to.infn.it}

\author[0000-0002-0998-4953]{Alberto Manfreda}
\affiliation{Istituto Nazionale di Fisica Nucleare, Sezione di Napoli, Strada Comunale Cinthia, 80126 Napoli, Italy}
\email{alberto.manfreda@pi.infn.it}

\author[0000-0003-3331-3794]{Fabio Muleri}
\affiliation{INAF Istituto di Astrofisica e Planetologia Spaziali, Via del Fosso del Cavaliere 100, 00133 Roma, Italy}
\email{fabio.muleri@inaf.it}

\author[0000-0002-9774-0560]{John Rankin}
\affiliation{INAF Istituto di Astrofisica e Planetologia Spaziali, Via del Fosso del Cavaliere 100, 00133 Roma, Italy}
\affiliation{INAF – OAB Merate, Via E. Bianchi 46, 23807 Merate, Italy}
\email{john.rankin@inaf.it }

\author[0000-0001-5676-6214]{Carmelo Sgrò}
\affiliation{Istituto Nazionale di Fisica Nucleare, Sezione di Pisa, Largo B. Pontecorvo 3, 56127 Pisa, Italy}
\email{carmelo.sgro@pi.infn.it}

\author[0000-0002-8665-0105]{Stefano Silvestri}
\affiliation{Istituto Nazionale di Fisica Nucleare, Sezione di Pisa, Largo B. Pontecorvo 3, 56127 Pisa, Italy}
\email{stefano.silvestri@pi.infn.it}

\author[0000-0002-5270-4240]{Martin C. Weisskopf}
\affiliation{NASA Marshall Space Flight Center, Huntsville, AL 35812, USA}
\email{martin.c.weisskopf@nasa.gov}

\begin{abstract}
We present a comprehensive characterization of the on-orbit imaging performance of the three telescopes on board the Imaging X-ray Polarimetry Explorer (IXPE). 
Each telescope comprises a Wolter-I mirror module assembly and a Gas Pixel Detector focal-plane detector unit (DU). 
We analyze data from point-like X-ray sources and fit a composite point spread function (PSF) model that we compare with ground calibrations.
We study the dependence of the PSF parameters and of the angular resolution, in terms of half-power diameter (HPD), on the time and source counting rate.
We find no significant secular evolution of PSF parameters or HPD over 30 months on orbit, with average HPDs of $26.1 \pm 0.5$ arcsec (Telescope 1), $32.1 \pm 0.5$ arcsec (Telescope 2), and $30.9 \pm 0.6$ arcsec (Telescope 3), and rate trends consistent with zero up to source counting rates of $\sim60$ cts s$^{-1}$ in the 2-3 keV energy band for all three telescopes. 
We set a 99\% C.L. upper limit of 4.4\% on the optics-induced polarization in the PSF halo, and find no measurable degradation of the polarization modulation factor in the wings versus the core due to mis-reconstructed photoelectron tracks.
IXPE’s imaging performance thus is consistent with the $\leq30$ arcsec observatory requirement with high stability, ensuring robust spatially resolved polarization measurements for the mission’s projected lifetime through 2030.
\end{abstract}

\keywords{Polarimetry, X-ray Telescopes}


\section{Introduction} 
X-ray polarimetry has emerged in recent years as a powerful diagnostic tool for probing high-energy astrophysical processes, ranging from the geometry of accretion disks around black holes to the magnetospheric structure of neutron stars, to the geometry of magnetic fields in non-plerionic supernova remnants and in pulsar wind nebulae\footnote[1]{See \url{https://www.mdpi.com/2075-4434/12/5} for a summary of the results obtained by IXPE.}
This breakthrough was enabled by gas-based photoelectric X-ray polarimeters, such as the Gas Pixel Detector \citep[GPD,][]{2001Costa,2006Bellazzini}, which allows the simultaneous measurement of the energy, time of arrival, and polarization state of the incoming photons.
The GPD also has the capability of determining the impact point of the events, thus providing astronomical imaging when put at the focus of an X-ray optic.
By resolving the incoming photon distribution on the focal plane, it is possible to isolate genuinely polarized emission from instrumental artifacts, subtract celestial and instrumental background in faint sources, and study the polarization as a function of the source morphology. 
This is crucial for the study of extended and diffuse sources such as supernova remnants and pulsar wind nebulae.
This synergy between imaging and polarimetry was first demonstrated in ground-based calibrations of a GPD coupled to a JET-X mirror module \citep{2014Fabiani}. 
The Imaging X-ray Polarimetry Explorer \citep[IXPE,][]{2022JWeisskopf,Soffitta2021}, launched in December 2021, is the first dedicated imaging-capable X-ray polarimetry mission. 
IXPE comprises three identical telescopes, each formed by a Wolter-I mirror module assembly (MMA) and a GPD in the focal plane.
For ease of identification, we  refer to each assembly as Telescope 1, 2, and 3. 
The MMAs were manufactured by NASA’s Marshall Space Flight Center \citep[][]{2022Ramsey} and each contain 24 nested mirror shells that concentrate and focus X-rays on the GPDs, which were developed and built in Italy \citep[][]{2021Baldini}. 
Together, this configuration achieves unprecedented sensitivity to linear polarization across the 2$–$8 keV band, while providing an angular resolution of the order of $\leq30$ arcsec. 
Rigorous end-to-end ground calibrations were performed at NASA’s Marshall Space Flight Center with the IXPE flight spare mirror module assembly (MMA-4) and the flight-spare detector (DU-FM1) \citep[][]{2025Ramsey}.
Despite this, on-orbit conditions inevitably alter the telescope point spread function (PSF) due to zero-gravity alignment shifts and thermal gradients different from those on ground. 
Further, a periodic (90 min) $\sim$ 1 arcmin wide boom-shift was observed after launch that, in the absence of a metrological system, is cured by applying a dynamic model and by an event-by-event re-centering of centroid positions calculated in time slices of the orbit, all performed in the processing pipeline. 
Consequently, obtaining an accurate sky-calibrated PSF for each telescope is essential.
The measured PSF includes contributions that make it deviate from the ideal -- i.e., azimuthally symmetric, smooth mirror surfaces and perfect centering coupled to a GPD that reconstructs impact points without bias -- optics/detector combination: inclined penetration and absorption of photons through the 1-cm thickness of the GPD absorption gap \citep[][]{2010Lazzarotto} that causes a few arcseconds of PSF degradation with respect to the intrinsic telescope angular resolution, the diffusion of the electrons when drifting in the gas, and track-reconstruction errors that produce wings outside the core.  
In fact, previous simulations \citep[][]{2013Soffitta} predicted that due to mis-reconstructed impact points, the wings of the PSF might exhibit a lower modulation factor - i.e. the detector response to 100\% polarized radiation - compared to the core.
Verifying this on orbit is important for constraining systematic effects in the polarization sensitivity of IXPE.
Understanding if and how the PSF components evolve over time and if and how they depend on source count rate is necessary both for precise flux and polarization measurements, and for estimating any optics-induced polarization (via grazing-incidence scattering). 
In this paper, we present the in-flight imaging performance of the three telescopes of IXPE. 
In Section \ref{sec:Methods} we describe our sample of point-like sources and the methodology for constructing radial profiles, fitting a composite PSF model, and deriving the angular resolution as the half-power diameter (HPD). 
In Section \ref{sec:Results} we examine the temporal and count-rate dependencies of the PSF fit parameters and HPD for each Telescope, assessing whether any trend persists before and after key operational milestones (optical realignment in mid-2022 and thermal set-point adjustment in late 2023).
We also study possible effects impacting the optics + detector system, such as the degree to which optics-scattering induces spurious polarization, and modulation factor decrease in the PSF wings. 
Finally, we summarize and discuss our findings in Section \ref{sec:Discussion}.

\section{Methods}
\label{sec:Methods}
We analyze level 2 data from 93 observations of X-ray point sources, both Galactic and extragalactic, observed by IXPE and available in the public archive.
For each observation, we process the data from three telescopes (Telescope 1, Telescope 2, and Telescope 3) independently. 
We chose sources with a count rate greater than 0.1 counts s$^{-1}$ in the 2$-$3 keV energy band, selected with the \texttt{ixpeobssim} \citep{2022Baldini} tool \texttt{xpselect}. 
This allows for a direct comparison of the in-flight parameters with those from the ground calibrations, which were obtained with a monochromatic beam at 2.3 keV, and allows for enough statistics to provide good fits of the PSF. 
To each source, we applied the instrumental background rejection algorithm as prescribed in \cite{2023DiMarco}. 
To characterize the imaging performance and angular resolution of the three IXPE telescopes, we first compute the position of the source on each detector using a Gaussian fit of the spatial distribution of the events.
We use the best-fit position to produce a weighted radial histogram by first calculating the radial distance of each event from the source centroid. 
We define concentric annuli with a specified bin width $\Delta r$ -- between 1 and 8 arcsec depending on the source counting rate -- and compute the counts in each annulus after weighting by the inverse of the annular area, i.e. $ \rm w(r)=1/(2\pi r \Delta r)$, in order to account for the geometrical area of each annulus.
We subtract from the radial histogram any residual background, estimated as the average of the count in the 10 outermost bins.
We use the resulting radial profile in subsequent modeling.
We describe the PSF with a function composed by the sum of a Gaussian core, a King profile to account for the extended wings, and an exponential term representing additional broad-scale structures that dominate the outer regions of the field of view:
\begin{equation}
\rm PSF(r) = W e^{-\frac{r^2}{2\sigma^2}} + N\Big[1 + \Big(\frac{r}{r_c}\Big)^2 \Big]^{-\eta} + M e^{-\frac{r}{r_0}} \quad.
\label{eq:PSF}
\end{equation}
In Eq. \ref{eq:PSF}, $\rm W$ and $\sigma$ are the Gaussian normalization and width; $\rm N$, $\rm r_c$, and $\eta$ are the King profile normalization, core radius, and wing slope; $\rm M$ and $\rm r_0$ are the exponential term normalization and scale.
With respect to the ground calibrations, this latter function replaces the power-law component as it requires one less parameter to fit with no losses on the PSF characterization.
The PSF defined in Equation \ref{eq:PSF} is analytically integrable in $\rm r'dr'$ and its integral profile is the encircled energy fraction (EEF):
\begin{equation}
\begin{split}
{\rm EEF}(r) &= \int_0^r {\rm PSF}(r')\,2\pi r'\,dr' = \\
&= 2\pi W\sigma^2\left(1-e^{-r^2/(2\sigma^2)}\right) 
+ \frac{\pi N r_c^2}{1-\eta}\left(\left[1+\left(\frac{r}{r_c}\right)^2\right]^{1-\eta}-1\right) + \\
&\quad + M r_0\left(1-e^{-r/r_0}\right).
\end{split}
\label{eq:eef}
\end{equation}
Taking the $r\to\infty$ limit (and noting that the King index satisfies $\eta>1$ for all fits reported here), the total flux of the source \textbf{is finite} and analytically characterized as
\begin{equation}
{\rm EEF}(\infty)=2\pi W\sigma^2 + \frac{\pi N r_c^2}{\eta-1} + M r_0.
\label{eq:eef_inft}
\end{equation}
We note that the exponential tail contributes negligibly to the encircled energy at small radii (e.g. within a few tens of arcseconds) but is important to reproduce observed surface brightness and total EEF at very large radii; its presence therefore affects EEF estimates at large apertures and the assessment of stray flux, while leaving the core calibration essentially unchanged.

We measure the angular resolution in terms of half power diameter (HPD), which is defined as twice the radius for which the integral in Eq. \ref{eq:eef} equals to $0.5$ and the uncertainty is estimated according to the binomial statistics as in \cite{2014Fabiani}. 
For each observation we compute the source position on the detector by fitting a two-dimensional Gaussian to the event spatial distribution; this best-fit position is then used as the center to build a background-subtracted, weighted, radial profile for PSF modeling within a fitting range of 0–300 arcsec.\\
The appropriate parameter bounds are set according to the expected behavior of the instrument \citep[see ][]{2025Ramsey}. 
The fit produced best-fit values and uncertainties for the parameters $\rm W$, $\rm \sigma$, $\rm N$, $\rm r_c$, $\rm \eta$, $\rm M$, and $\rm r_0$.
These parameters fully describe the PSF for each source and each Telescope, resulting in the EEF described in Eq. \ref{eq:eef} that we integrate to estimate the HPD.
We note that individual fitted parameter values often scatter beyond the formal fit uncertainties. 
This extra scatter arises from small azimuthal PSF asymmetries (not captured by the azimuthally symmetric model), residual centroiding uncertainty after the boom motion correction performed in the pipeline, imperfect background estimation for low-rate observations, and intrinsic source variability for a subset of targets. 
To ensure robust epoch average values and uncertainties we therefore report epoch averages as the arithmetic mean and quote the epoch uncertainty as the rms of the measurements. 
We verified that removing low-rate observations or sub-selecting epochs does not change the inferred trends within the quoted uncertainties.

\section{Results} 
\label{sec:Results}
\begin{figure}[htbp]
\begin{center}
\includegraphics[width=0.7\linewidth]{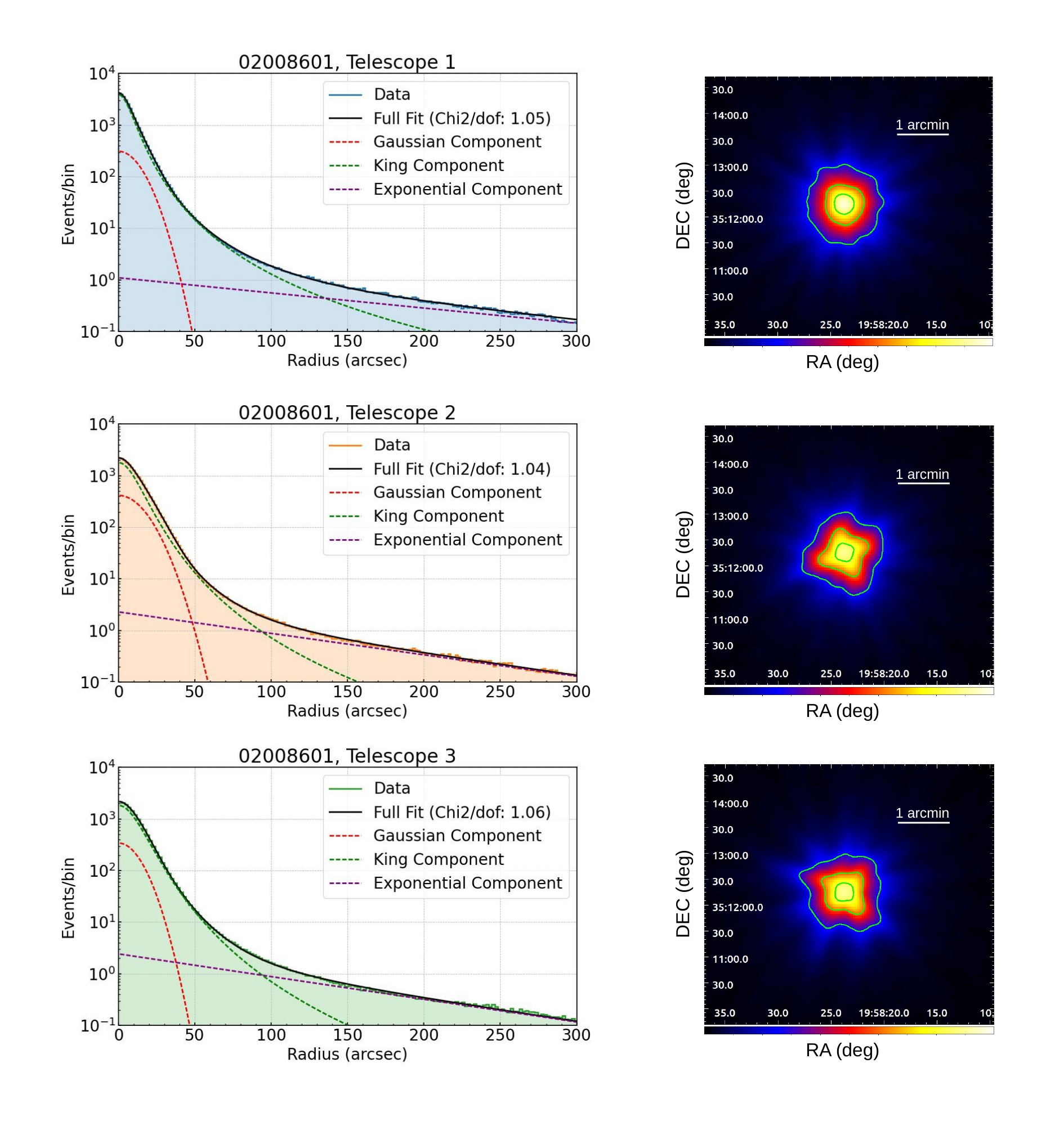}
\caption{ 
From top to bottom, best fit of the radial profiles and source image in log-scale of the three telescopes for an exemplary observation, obsid 2008601 (Cyg X-1).
In the right column, the green contours on the source images enclose 50\%, 90\%, and 99\% of the total counts.}
\label{fig:psf_v3}
\end{center}
\end{figure}

\begin{figure}[htbp]
\begin{center}
\includegraphics[width=0.4\linewidth]{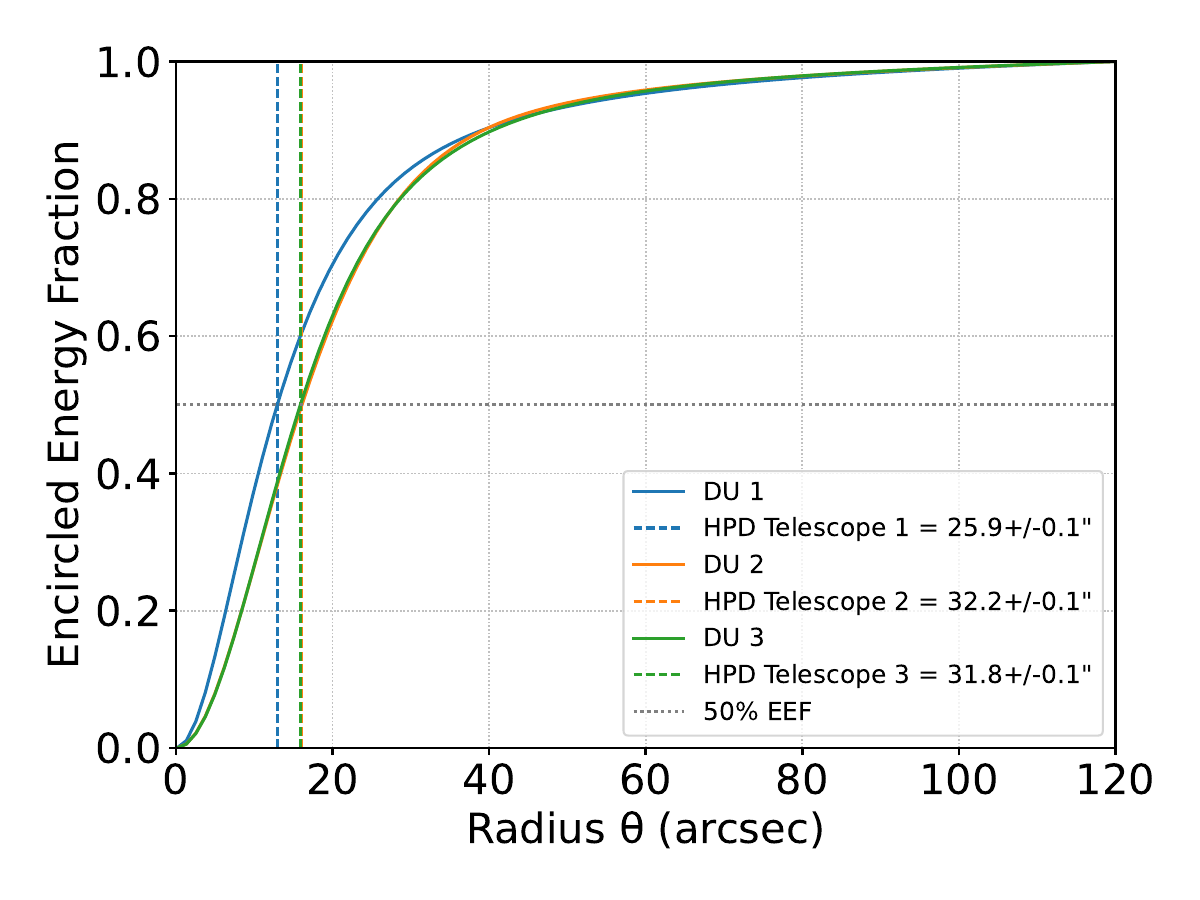}
\caption{ 
Encircled energy fraction (EEF) for the three telescopes resulting from the integration of the fits shown in Fig. \ref{fig:psf_v3} as a function of the angular distance from the source position for obsid 2008601 (Cyg X-1) in the 2-3 keV energy band.
The vertical dashed lines mark the the radius enclosing 50\% of the EEF, i.e. half of the Half Power Diameter (HPD).}
\label{fig:02008601_EEF_2_3_keV}
\end{center}
\end{figure}

We report below for each Telescope the results of the fit of the IXPE data with Eq. \ref{eq:PSF}.
The detailed fit results for Telescope 1, Telescope 2, and Telescope 3 across all 93 observations are tabulated in the Appendix in Table \ref{tab:Telescope_1_fit_results}, \ref{tab:Telescope_2_fit_results}, \ref{tab:Telescope_3_fit_results}.
As an example, we show the fit for each telescope for ObsID 02008601, the bright accreting black hole Cygnus X-1 in Fig. \ref{fig:psf_v3}. 
The King core dominates within $\sim10$ arcsec -- with significant contribution from the Gaussian core --, the King exponential term accounts for the $10–100$ arcsec wings, and the exponential tail describes emission beyond $\sim100$ arcsec. 
We show in Fig. \ref{fig:02008601_EEF_2_3_keV} the EEF as a function of the angular separation from the source position for ObsID 02008601 and the resulting HPD for each Telescope.
We note that, although our composite PSF model assumes azimuthal symmetry by construction, in practice Telescope2 and 3 exhibit mild departures from circular symmetry (right column of Fig. \ref{fig:psf_v3}), yet the radially averaged profiles remain well fit by the symmetric model.
This indicates that the asymmetries are sufficiently small and distributed around the full 360° that their contributions effectively cancel in the radial average.
We do observe that the best‑fit PSF parameters for Telescope 2 and 3 deviate more from their ground‑calibration values than Telescope 1, consistent with slightly larger unmodeled structure in these units.
\subsection{Time and rate behavior of fit parameters}
We fit the results with a linear regression to identify any trend.
We also split the data into three epochs based on two mission events with possible repercussions on the angular resolution that occurred after the beginning of the observations of IXPE in January 2022: a realignment of the optical modules using the on-board actuator system, the so-called tip-tilt-rotate mechanism, that occurred in June 2022 at mission elapsed time (MET) 1.713e8 s, and an optics temperature set-point change that was done in September 2023 at MET 2.140e8 s to extend the spacecraft battery life.
In the following, we define the epochs bookmarked by these two events as T1, T2, and T3.
The slope of the linear regression for each time period is calculated with respect to a $\rm T_{mid}$ corresponding to the middle of the epoch: 1.665e8 s for T1, 1.925e8 s for T2, and 2.362e8 s for T3.
For ease of interpretation and visualization, in the fit results we show the slope of the fit in units of arcsec day$^{-1}$.
For each time interval, we provide the best-fit line and the average parameter value as 
\begin{equation}
     \rm value(t) = slope \ [arcsec \ day^{-1}] *(t - T_{mid}) + offset [arcsec]  \quad .
\end{equation}
The uncertainty on the average is given by the rms of the points. 
The trend versus source counting rate is instead calculated for all time periods as
\begin{equation}
     \rm value(rate) = slope \ [arcsec \ counts^{-1} \ s] *(rate) + offset [arcsec]  \quad .
\end{equation}
An exception is represented by the parameter $r_0$, as a simple linear trend is not optimal to describe the rate trend.
We found that the best fit is provided by a function of the form
\begin{equation}
    \rm r_0(rate) = A + B*rate^{-C} \quad .
\end{equation}

\subsubsection{Gaussian width $\sigma$}
Ground calibration at 2.3 keV yielded for the spare Telescope $\sigma^{ground} =12.3 \pm 0.1$ arcsec.
On orbit, in all time periods, the time slopes are consistent with zero, indicating that there is no significant secular degradation of the Gaussian core width. 
Fitting $\sigma$ against the source rate yields mild positive correlations for high source rates only for Telescope 1 and Telescope 2.
The fit results are shown in Fig.\ref{fig:sigma_fit} and tabulated in Table \ref{tab:sigma_fit}.\\
\begin{minipage}[t]{1.\textwidth}
\captionof{figure}{Fit of the Gaussian width $\sigma$ as a function of time (left panel) and source 2-3 keV rate (right panel) for all 3 detectors on-board IXPE. 
The red vertical line marks the epoch of optics realignment, the blue vertical line marks the epoch of the temperature set-point change, and the pink shaded line the ground-calibration value and its uncertainties.
The shaded areas cover the 1 standard deviation uncertainty of the linear regression.}  
\label{fig:sigma_fit}
\includegraphics[width=\linewidth]{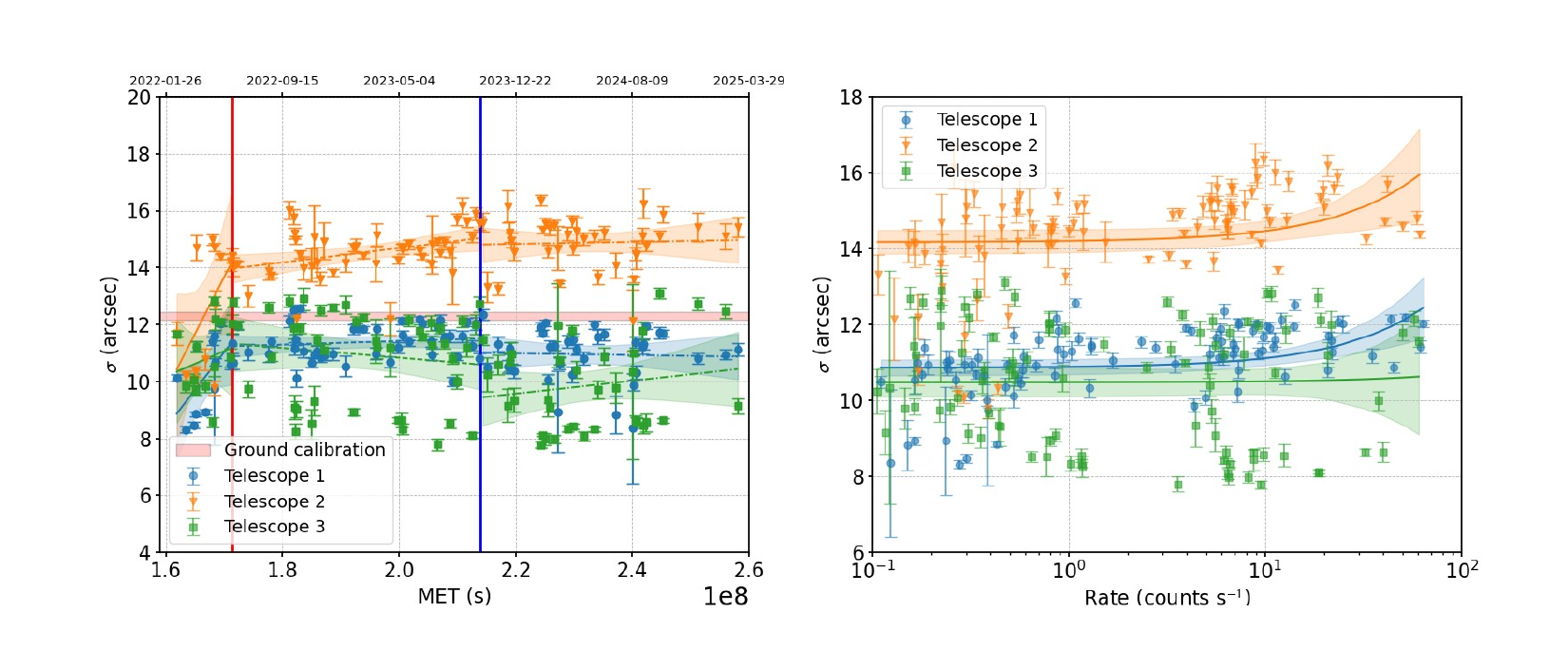} 
\captionof{table}{Fit results for the time and rate trend of the Gaussian width $\sigma$ for each Telescope in the three time period considered.}
\label{tab:sigma_fit}
\centering
\begin{tabular}{l|llll|ll}
                   & \multicolumn{4}{c}{\textbf{Time trend}}                            & \multicolumn{2}{c}{\textbf{Rate trend}}                             \\
\hline
                   & \textbf{Time}     & \textbf{Offset}      & \textbf{Slope}               & \textbf{Average}    & \textbf{Offset}                     & \textbf{Slope}                         \\
\textbf{Telescope}                 & \textbf{interval} & \textbf{[arcsec]}    & \textbf{[arcsec day$^{-1}$]} & \textbf{[arcsec]}   & \textbf{[arcsec]}                   & \textbf{[arcsec counts$^{-1}$ s]}      \\
\hline
\multirow{3}{*}{1} & T1            & 10.1 ± 0.3                                                    & 0.023 ± 0.011                                                    & 10.2 ± 1.1 & \multirow{3}{*}{10.9 ± 0.1}                                   & \multirow{3}{*}{0.025 ± 0.007}                                   \\
                   & T2            & 11.4 ± 0.1                                                    & 0.0005 ± 0.0006                                                  & 11.4 ± 0.3 &                                                               &                                                                  \\
                   & T3            & 10.9 ± 0.2                                                    & -0.0003 ± 0.0012                                                 & 10.9 ± 0.5 &                                                               &                                                                  \\
\hline
\multirow{3}{*}{2} & T1            & 12.3 ± 0.6                                                    & 0.034 ± 0.019                                                    & 12.4 ± 1.7 & \multirow{3}{*}{14.2 ± 0.2}                                   & \multirow{3}{*}{0.029 ± 0.011}                                   \\
                   & T2            & 14.5 ± 0.1                                                    & 0.002 ± 0.001                                                    & 14.5 ± 0.4 &                                                               &                                                                  \\
                   & T3            & 14.9 ± 0.2                                                    & 0.0003 ± 0.0012                                                  & 14.9 ± 0.5 &                                                               &                                                                  \\
\hline
\multirow{3}{*}{3} & T1            & 10.8 ± 0.4                                                    & 0.008 ± 0.013                                                    & 10.8 ± 1.0 & \multirow{3}{*}{10.5 ± 0.2}                                   & \multirow{3}{*}{0.003 ± 0.014}                                   \\
                   & T2            & 11.0 ± 0.2                                                    & -0.002 ± 0.002                                                   & 10.9 ± 0.7 &                                                               &                                                                  \\
                   & T3            & 9.9 ± 0.3                                                     & 0.002 ± 0.002                                                    & 9.8 ± 0.8  &                                                               &                                                                 \\
                   \hline
\end{tabular}
\end{minipage}
%
\subsubsection{King core radius $r_c$}
Ground calibration for the spare Telescope gave $r_{c}^{ground} =9.7 \pm 0.2$ arcsec.
On-orbit fits do not produce significant time slopes.
The correlations with rate are weak: Telescope 2 shows a marginal negative trend, while Telescope 1 and Telescope 3 remain rate-independent.
The fit results are shown in Fig.\ref{fig:r_c_fit} and tabulated in Table \ref{tab:r_c_fit}.\\
\begin{minipage}[t]{1.\textwidth}
\captionof{figure}{Fit of the King core radius $r_c$ as a function of time (left panel) and source 2-3 keV rate (right panel) for all 3 detectors on-board IXPE. 
The red vertical line marks the epoch of optics realignment, the blue vertical line marks the epoch of the temperature set-point change, and the pink shaded line the ground-calibration value and its uncertainties.
The shaded areas cover the 1 standard deviation uncertainty of the linear regression.}  
\label{fig:r_c_fit}
\includegraphics[width=\linewidth]{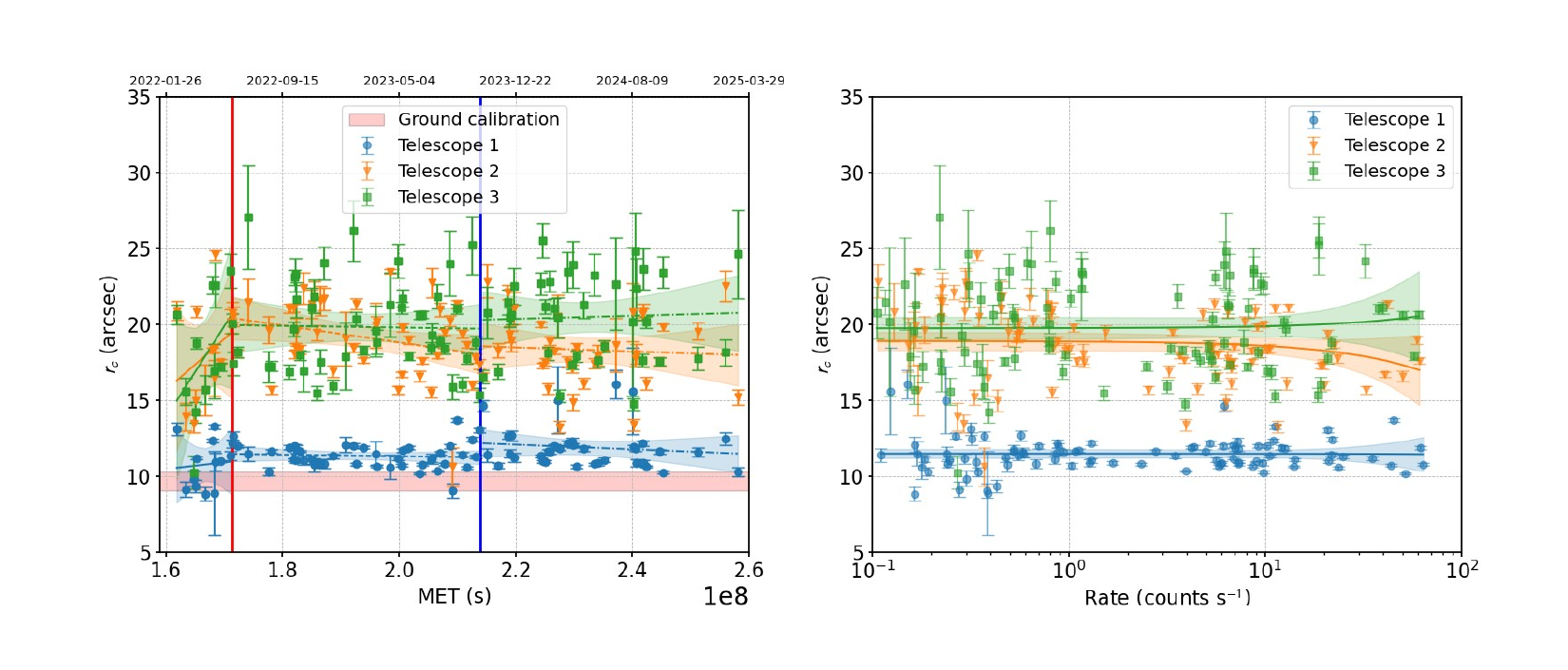} 
\captionof{table}{Fit results for the time and rate trend of the King core radius $r_c$ for each Telescope in the three time period considered.}
\label{tab:r_c_fit}
\centering
\begin{tabular}{l|llll|ll}
                   & \multicolumn{4}{c}{\textbf{Time trend}}                                                & \multicolumn{2}{c}{\textbf{Rate trend}}         \\
\hline
                   & \textbf{Time}     & \textbf{Offset}   & \textbf{Slope}              & \textbf{Average} & \textbf{Offset}               & \textbf{Slope}           \\
\textbf{Telescope}        & \textbf{interval} & \textbf{[arcsec]} & \textbf{[arcsec day$^{-1}$]}& \textbf{[arcsec]}& \textbf{[arcsec]}             & \textbf{[arcsec counts$^{-1}$ s]}\\
\hline
\multirow{3}{*}{1} & T1                & 10.8 ± 0.5        & 0.004 ± 0.016               & 10.8 ± 1.3       & \multirow{3}{*}{11.5 ± 0.1}   & \multirow{3}{*}{0.001 ± 0.010}\\
                   & T2                & 11.4 ± 0.1        & -0.0005 ± 0.0008            & 11.4 ± 0.4       &                               &                               \\
                   & T3                & 11.8 ± 0.2        & -0.001 ± 0.002              & 11.9 ± 0.7       &                               &         \\
\hline
\multirow{3}{*}{2} & T1                & 17.9 ± 1.0        & 0.029 ± 0.032               & 17.9 ± 2.7       & \multirow{3}{*}{18.9 ± 0.3}   & \multirow{3}{*}{-0.031 ± 0.021} \\
                   & T2                & 19.2 ± 0.3        & -0.005 ± 0.002              & 19.1 ± 1.1       &                               &       \\
                   & T3                & 18.3 ± 0.4        & -0.001 ± 0.003              & 18.3 ± 1.1       &                               &      \\
\hline
\multirow{3}{*}{3} & T1                & 17.8 ± 1.1        & 0.052 ± 0.037               & 17.9 ± 3.2       & \multirow{3}{*}{19.7 ± 0.4}   & \multirow{3}{*}{0.013 ± 0.027} \\
                   & T2                & 19.8 ± 0.4        & -0.0005 ± 0.0030            & 19.8 ± 1.3       &                               &                           \\
                   & T3                & 20.5 ± 0.5        & 0.001 ± 0.004               & 20.5 ± 1.4       &                               &          \\
\hline
\end{tabular}
\end{minipage}
\subsubsection{King wing slope $\eta$}
Ground calibration for the spare Telescope at 2.3 keV yielded $\eta^{ground} = 1.95 \pm 0.03$. 
Time dependencies are negligible for all telescopes and weak rate trends appear only for Telescope 3.
The fit results are shown in Fig.\ref{fig:eta_fit} and tabulated in Table \ref{tab:eta_fit}.\\
\begin{minipage}[t]{1\textwidth}
\centering
\captionof{figure}{Fit of the King wing slope $\eta$ as a function of time (left panel) and source 2-3 keV rate (right panel) for all 3 detectors on-board IXPE. 
The red vertical line marks the epoch of optics realignment, the blue vertical line marks the epoch of the temperature set-point change, and the pink shaded line the ground-calibration value and its uncertainties.
The shaded areas cover the 1 standard deviation uncertainty of the linear regression.}  
\label{fig:eta_fit}
\includegraphics[width=\linewidth]{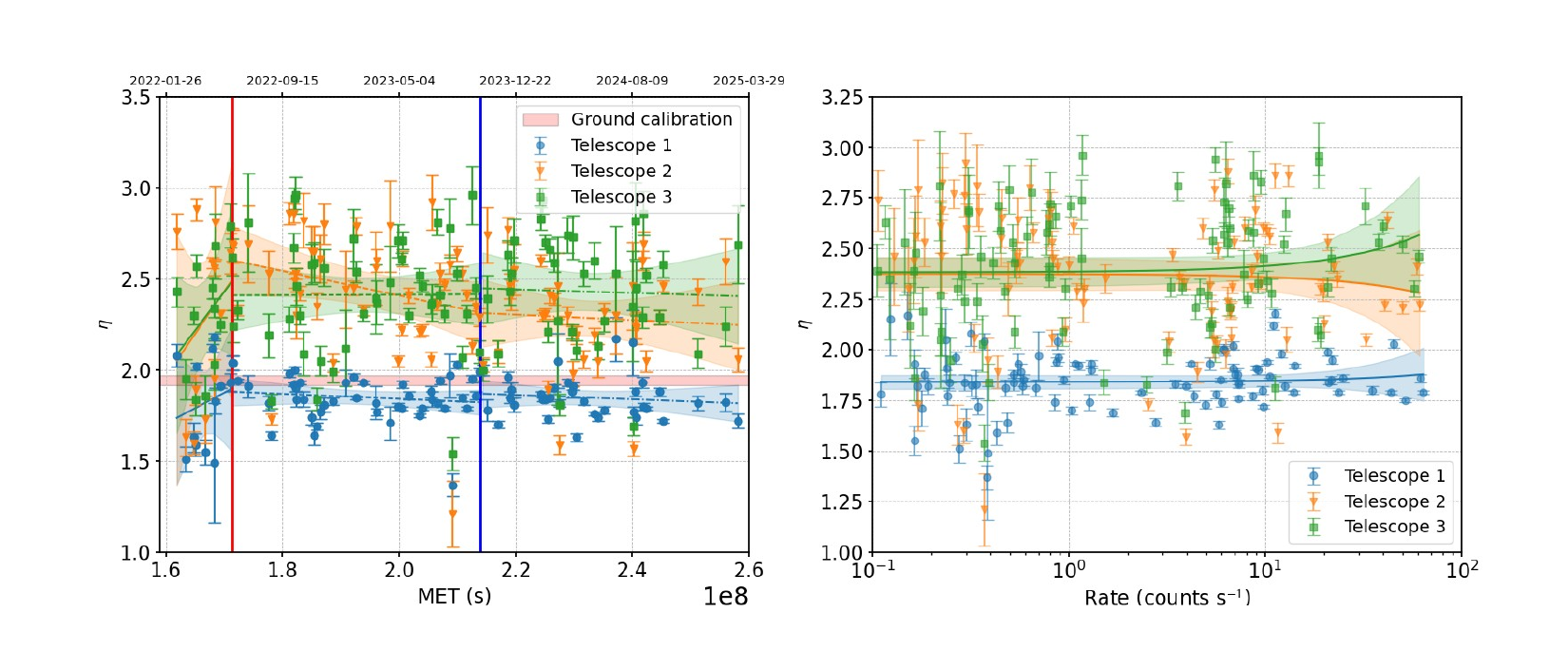} 
\captionof{table}{Fit results for the time and rate trend of the King wing slope $\eta$ for each Telescope in the three time period considered.}
\label{tab:eta_fit}
\centering
\begin{tabular}{l|llll|ll}
                   & \multicolumn{4}{c}{\textbf{Time trend}}                                      & \multicolumn{2}{c}{\textbf{Rate trend}}             \\
\hline
                   & \textbf{Time}     & \textbf{Offset}& \textbf{Slope}       & \textbf{Average} & \textbf{Offset}              & \textbf{Slope}  \\
\textbf{Telescope}        & \textbf{interval} &                & \textbf{[day$^{-1}$]}&                  &                              & \textbf{[counts$^{-1}$ s]} \\
\hline
\multirow{3}{*}{1} & T1                & 1.82 ± 0.08    & 0.002 ± 0.003        & 1.8 ± 0.2        & \multirow{3}{*}{1.84 ± 0.02} & \multirow{3}{*}{0.001 ± 0.001} \\
                   & T2                & 1.85 ± 0.02    & -0.0001 ± 0.0001     & 1.8 ± 0.1        &                              &      \\
                   & T3                & 1.84 ± 0.02    & -0.0001 ± 0.0002     & 1.8 ± 0.1        &                              &      \\
\hline
\multirow{3}{*}{2} & T1                & 2.27 ± 0.14    & 0.004 ± 0.005        & 2.3 ± 0.4        & \multirow{3}{*}{2.37 ± 0.04} & \multirow{3}{*}{-0.002 ± 0.003}\\
                   & T2                & 2.46 ± 0.05    & -0.0006 ± 0.0003     & 2.4 ± 0.1        &                              &           \\
                   & T3                & 2.28 ± 0.05    & -0.0001 ± 0.0004     & 2.3 ± 0.1        &                              &        \\
\hline
\multirow{3}{*}{3} & T1                & 2.28 ± 0.09    & 0.004± 0.003         & 2.3 ± 0.3        & \multirow{3}{*}{2.38 ± 0.04} & \multirow{3}{*}{0.003 ± 0.003}\\
                   & T2                & 2.41 ± 0.05    & 0.0000 ± 0.0003      & 2.4 ± 0.1        &                              &            \\
                   & T3                & 2.43 ± 0.05    & -0.0001 ± 0.0004     & 2.4 ± 0.1        &                              &           \\
\hline
\end{tabular}
\end{minipage}
\subsubsection{Exponential scale $r_0$}
Because the ground calibration data for the spare Telescope were fitted with a power law term for the PSF wings, in order to compare it with the exponential used for the in-flight data, we equate the integrals of the two functions for the ground data and solve for the equivalent $\rm r_0$ parameter, establishing a ground calibration value of $r_0^{ground} = 113.6 \pm 9.2$ arcsec.
The time trend is highly affected by the spread of the values, because of the marginal rate correlation, especially below 1 count s$^{-1}$.
For this reason, we consider for the time trend study, only the observations with counting rate above 1 count s$^{-1}$.
We find that for bright sources, both the time and rate trends appear to be compatible with the ground calibration measurement, indicating no strong rate-driven tail broadening and time stability. 
The fit results are shown in Fig.~\ref{fig:r0_fit} and tabulated in Table \ref{tab:r0_fit}.
%
\begin{minipage}[t]{1.\textwidth}
\centering
\captionof{figure}{Fit of the exponential scale $r_0$ as a function of time (left panel) and source 2-3 keV rate (right panel) for all 3 detectors on-board IXPE. 
The red vertical line marks the epoch of optics realignment, the blue vertical line marks the epoch of the temperature set-point change, and the pink shaded line the ground-calibration value and its uncertainties.
The shaded areas cover the 1 standard deviation uncertainty of the linear regression.
The shaded data points in the left panel are the ones from sources with 2-3 keV counting rate $<1$ count s$^{-1}$ that are excluded from the time trend fitting.}  
\label{fig:r0_fit}
\includegraphics[width=\linewidth]{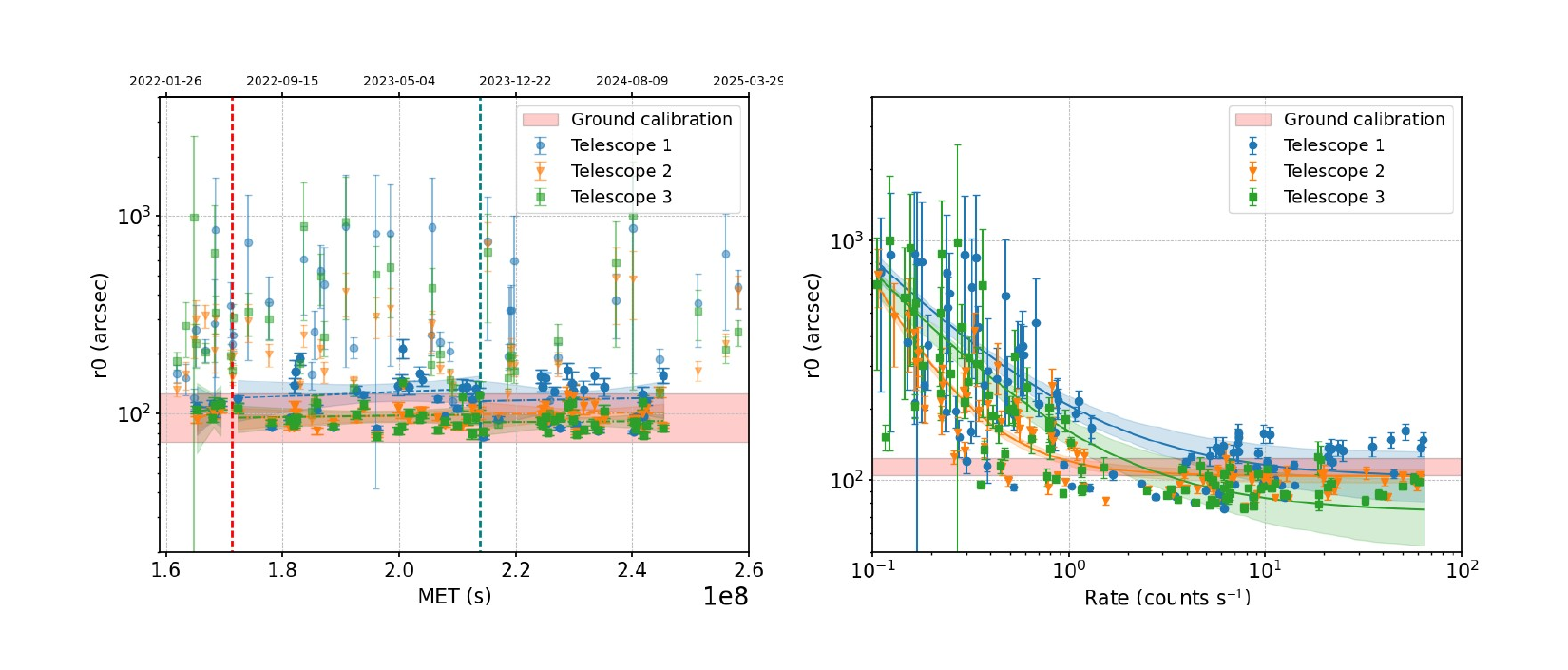} 
\captionof{table}{Fit results for the time and rate trend of the exponential scale $r_0$ for each Telescope in the three time period considered.}
\label{tab:r0_fit}
\centering
\begin{tabular}{l|llll|lll}
                   & \multicolumn{4}{c}{\textbf{Time trend}}                                                            & \multicolumn{3}{c}{\textbf{Rate trend}} \\
\hline
                   & \textbf{Time}      & \textbf{Offset}   & \textbf{Slope}               & \textbf{Average}  & \textbf{A}                    & \textbf{B}                    & \textbf{C}       \\
\textbf{Telescope}        &  \textbf{interval} & \textbf{[arcsec]} & \textbf{[arcsec day$^{-1}$]} & \textbf{[arcsec]} & \textbf{[arcsec]}             &  \textbf{[arcsec]}            &                     \\
\hline
\multirow{3}{*}{1} & T1                 & 108.4 ± 3.7       & 0.02 ± 0.21                  & 108.5 ± 7.6       & \multirow{3}{*}{101.4 ± 28.2} & \multirow{3}{*}{103.7 ± 36.0} & \multirow{3}{*}{0.86 ± 0.18} \\
                   & T2                 & 127.1 ± 6.5       & 0.029 ± 0.043                & 128.4 ± 18.0      &                               &                               &     \\
                   & T3                 & 118.0 ± 6.4       & 0.012 ± 0.067                & 118.3 ± 17.9      &                               &                               &     \\
\hline
\multirow{3}{*}{2} & T1                 & 99.0 ± 1.7        & 0.215 ± 0.091                & 100.0 ± 6.5       & \multirow{3}{*}{103.8 ± 6.7}  & \multirow{3}{*}{16.2 ± 4.5}   & \multirow{3}{*}{1.56 ± 0.14} \\
                   & T2                 & 97.0± 2.7         & 0.006 ± 0.018                & 97.2 ± 7.5        &                               &                               &   \\
                   & T3                 & 100.3 ± 3.0       & 0.007 ± 0.031                & 100.5 ± 8.2       &                               &                               &      \\
\hline
\multirow{3}{*}{3} & T1                 & 105.5 ± 5.0       & 0.111 ± 0.275                & 106.0 ± 10.6      & \multirow{3}{*}{73.2 ± 26.2}  & \multirow{3}{*}{86.5 ± 31.8}  & \multirow{3}{*}{0.89 ± 0.18} \\
                   & T2                 & 97.2 ± 3.4        & 0.007 ± 0.022                & 97.6 ± 9.2        &                               &                               &     \\
                   & T3                 & 91.4 ± 2.5        & 0.003 ± 0.026                & 91.5 ± 6.7        &                               &                               &      \\    
\hline
\end{tabular}
\end{minipage}
\subsection{Half-Power Diameter (HPD)}
Ground HPD at 2.3 keV for the spare telescope was $22.2\pm0.1$ arcsec. 
In Fig. \ref{fig:HPDvsTime} we show the HPD values for each Telescope as a function of time and source rate in the 2$-$3 keV energy band.
On-orbit all time slopes are statistically consistent with zero, indicating no secular HPD degradation.
We found no significant HPD–rate correlation for all telescopes.
The fit results are tabulated in Table \ref{tab:hpd_fit}.
\begin{minipage}[t]{1.\textwidth}
\centering
\captionof{figure}{Fit of the HPD as a function of time (left panel) and source 2-3 keV rate (right panel) for all 3 detectors on-board IXPE. 
The red vertical line marks the epoch of optics realignment, the blue vertical line marks the epoch of the temperature set-point change, and the pink shaded line the ground-calibration value and its uncertainties.
The shaded areas cover the 1 standard deviation uncertainty of the linear regression.}  
\label{fig:HPDvsTime}
\includegraphics[width=\linewidth]{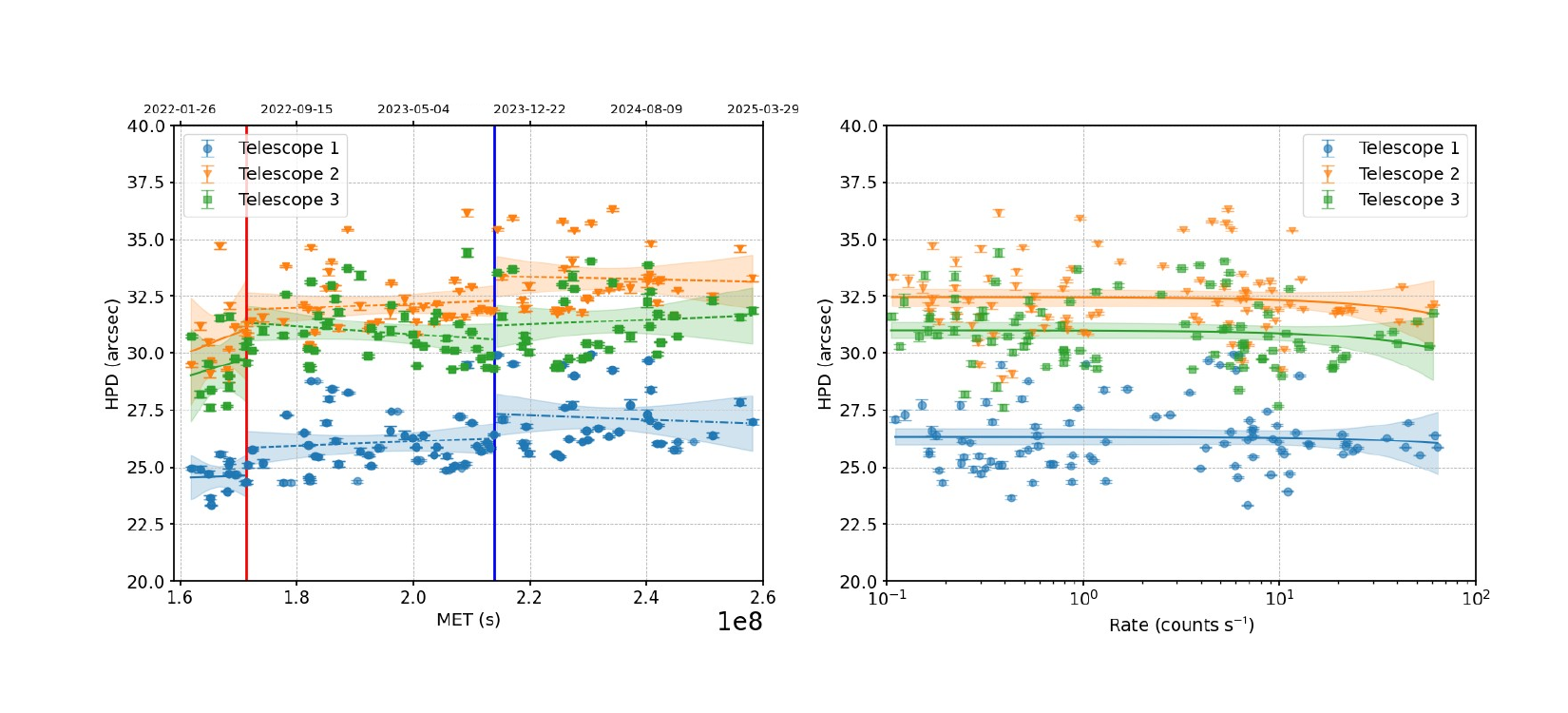} 
\captionof{table}{Fit results for the time and rate trend of the HPD for each Telescope in the three time period considered.}
\label{tab:hpd_fit}
\centering
\begin{tabular}{l|llll|ll}
                   & \multicolumn{4}{c}{\textbf{Time trend}}                            & \multicolumn{2}{c}{\textbf{Rate trend}}                             \\
\hline
                   & \textbf{Time}     & \textbf{Offset}      & \textbf{Slope}               & \textbf{Average}    & \textbf{Offset}                     & \textbf{Slope}                         \\
\textbf{Telescope}                 & \textbf{interval} & \textbf{[arcsec]}    & \textbf{[arcsec day$^{-1}$]} & \textbf{[arcsec]}   & \textbf{[arcsec]}                   & \textbf{[arcsec counts$^{-1}$ s]}      \\
\hline
\multirow{3}{*}{1} & T1       & 24.6 ± 0.2  & 0.0007 ± 0.0068     & 24.6 ± 0.6 & \multirow{3}{*}{26.3 ± 0.2}& \multirow{3}{*}{0.005 ± 0.012}\\
                   & T2       & 26.0 ± 0.2  & 0.0009 ± 0.0012     & 26.1 ± 0.5 &                            &                               \\
                   & T3       & 27.1 ± 0.2  & -0.0008 ± 0.0018    & 27.2 ± 0.7 &                            &                               \\
\hline
\multirow{3}{*}{2} & T1       & 30.6 ± 0.5  & 0.0088 ± 0.0165     & 30.6 ± 1.4 & \multirow{3}{*}{32.5 ± 0.2}& \multirow{3}{*}{-0.012 ± 0.013}\\
                   & T2       & 32.1 ± 0.2  & 0.0008 ± 0.0012     & 32.1 ± 0.5 &                            &                                \\
                   & T3       & 33.2 ± 0.2  & -0.0005 ± 0.0018    & 33.3 ± 0.7 &                            &                                \\
\hline
\multirow{3}{*}{3} & T1       & 29.4 ± 0.4  & 0.0067 ± 0.0142     & 29.4 ± 1.2 & \multirow{3}{*}{31.0 ± 0.2}& \multirow{3}{*}{-0.013 ± 0.013}\\
                   & T2       & 31.0 ± 0.2  & -0.0015 ± 0.0013    & 30.9 ± 0.6 &                            &                                \\
                   & T3       & 31.4 ± 0.3  & 0.0009 ± 0.0019     & 31.4 ± 0.7 &                            &                                \\
\hline
\end{tabular}
\end{minipage}

\subsection{Optics + GPD system \textbf{instrumental effects}}
We use IXPE observation to test two instrumental effects that can \textbf{impact the behavior} of the optics+instrument system in the events outside the PSF core.

\subsubsection{Optics-induced polarization}
Reflection of grazing incidence on the smooth Wolter I mirrors employed by the IXPE telescopes is described by the Fresnel equations and by the rules of optical geometry, but they assume "perfect" reflecting surfaces.
In principle, no net polarization is introduced, because azimuthal contributions cancel out. 
In practice, small-scale surface roughness scatters photons into a halo outside the PSF core, potentially breaking that symmetry.
Indeed, we already noted in Section \ref{sec:Results} that the images of especially Telescope 2 and 3 exhibit modest asymmetries from the ideal -- i.e., azimuthally symmetric, smooth mirror surfaces and perfect centering -- shape.
Under this interpretation, we could expect some effect on the mirror reflectivity: scattering removes some higher energy photons, with the amount of scattering being proportional to the square of the photon energy, but in the current modeling it is always assumed that the roughness is geometric and the reflection is still computed with this geometry and Fresnel equations.
From another point of view, the photons in the scattering halo have been scattered at angles different from the nominal one.
This means that the scattered photons are focused at a point different from those of the core.
This also means that the sum of different angular contributions that effectively cancels the polarization induced by the optics is no more active.
In order to estimate the typical magnitude of this scattering-induced polarization, we consider the reflection coefficients for parallel ($r_p$) and perpendicular ($r_s$) polarization given by the Fresnel equations:
\begin{equation}
    r_p = \Big(\frac{E_R}{E_i}\Big)_p = \frac{n^2 \sin \alpha - \sqrt{n^2 - \cos^2 \alpha}}{n^2 \sin \alpha + \sqrt{n^2 - \cos^2 \alpha}}  \quad ;
    \label{eq:r_p}
\end{equation}
\begin{equation}
    r_s = \Big(\frac{E_R}{E_i}\Big)_s = \frac{\sin \alpha - \sqrt{n^2 - \cos^2 \alpha}}{\sin \alpha + \sqrt{n^2 - \cos^2 \alpha}} \quad .
    \label{eq:r_s}
\end{equation}
Here $n = 1 - \delta - i\beta$ is the complex index of refraction of the mirror - with $\delta$ describing the phase change and $\beta$ accounting for the absorption - and $E_R/E_i$ denotes the ratio of the amplitudes of the reflected and incident electric fields and $\alpha$ is the grazing angle of incidence as measured from the interface plane - i.e. the local surface plane at the point where the X-ray photon encounters the mirror.
Assuming 2 keV photons grazing on the most external shell of the IXPE telescopes, $\alpha=0.49°$, Eq.s \ref{eq:r_p} and \ref{eq:r_s} give, respectively $r_p = 0.8926$ and $r_s = 0.892524$, so that the polarization induced by two reflections will be $2(r_p-r_s)/(r_p+r_s) \sim 10^{-4}$.
At 7 keV, the effect would be about half of this value.
However, each angular segment of the shell will produce a polarization with the angle tangent at that point. 
All these azimuthal contributions will sum up in the PSF core and the total effect will be zero (or better, a polarization dilution of the order of $\sim10^{-4}$).
Still, the possible effect of the optics on the polarization is one of the most frequently asked questions about IXPE so, in order to empirically measure an upper bound to this effect, we consider an IXPE observation of a bright, unpolarized source: the weakly magnetized neutron star X-ray binary GS 1826-238 (obsid 1002801, \citet{2023Capitanio}).
We consider an annular region centered on the source with inner radius of 2 arcminutes and outer radius of 4.5 arcminutes in the 2$–$6 keV energy band  - thus removing the PSF core and excluding the background-rich detector borders and energies, and maximizing the statistics at the same time.
Considering all 3 telescopes, we derive a 99\% C.L. upper limit of 4.4\% polarization in the halo. 
To improve the significance of the possible tangential effect, we use the \texttt{ixpeobssim} tool \texttt{stokesalign} that aligns the Stokes parameters to a given polarization model on an event-by-event basis.
We find a non significant measurment PD of $1.7\pm1.4\%$, that is consistent with the MDP upper limit, and an unconstrained polarization angle.
We also verified using the \texttt{leakagelib}\citep[][]{2024Dinsmore} that in the halo there is no significant contamination of radial polarization of the source due to leakage effects.
Thus, at least 95.6 \% of grazing‐reflection photons conform to Fresnel behavior, whereas any optics‐induced polarization must be below 4.4 \% in the outer halo region.
To further put into context the weight of this potential effect, the source flux in the halo amounts to $\sim4.5\%$ of the total source flux.

\subsubsection{Impact-point reconstruction errors}
Monte Carlo simulations \citep{2013Soffitta} showed that the distribution of the photoelectron charges for some kind of trajectories provides blurred collected tracks with the skewness giving the ’wrong’ sign, leading to the derivation of the incorrect track-end point and the onset of "wings" outside of the PSF core. 
This happens because the photoelectron-track reconstruction algorithm calculates the skewness to determine the end point of the track containing the Bragg peak, which is the end of the photoelectron evolution.
The impact on the polarization sensitivity is that the core has a larger polarization sensitivity with respect to the wings by a factor of almost two.
To test this, we consider an IXPE observation of the bright, highly polarized and point-like accreting black hole Cyg X-3.
We select the core, defined as a HPD-wide circle centered on the source, and the wings, defined as the annulus between the HPD and 4 times the HPD - amounting to an EEF$\geq99\%$ and corresponding to the region in which the King + Exponential tails dominate over the Gaussian core - in the 5 - 7 keV energy band.
This allows for a comparison with the 6 keV simulated beam by \citet{2013Soffitta}.
We find that the Core PD is $21.4\pm1.3$, while the wings PD is $22.3\pm2.1\%$.
Hence, within uncertainties, the wings exhibit no statistically significant degradation of modulation compared to the core.
The lower modulation expected from Monte Carlo simulation found with a narrow beam is strongly masked by the fact that the real psf of a telescope is wider and includes effects of finite telescope resolution, scattering, as described above, and broadening resulting from inclined penetration of photons in the gas.

\section{Discussion and Conclusions} \label{sec:Discussion}

We performed a systematic study of the on-orbit PSF characteristics of the IXPE telescopes considering data from 93 point-like sources. 
We performed a systematic fit of the IXPE PSF with a composite model and studied the time and rate evolution of the fit parameters.
We find most parameters to significantly differ from the values measured during the ground calibration.
These variations can be attributed to the relaxation of the mirror mounting.
Indeed, the source image for each detector (Fig. \ref{fig:psf_v3}), especially for Telescope 1 and Telescope 3, show clear asymmetries with respect to an ideal circular PSF, whereas Telescope 1 is more similar to the near circular PSF seen in ground calibrations of the spare telescope \citep[][]{2025Ramsey}. 
As it is clear also from Fig. \ref{fig:02008601_EEF_2_3_keV}, the PSF cores differ due to differences in geometry resulting from figure errors in the shells and distortions due to mounting, while the wings remain similar as the surfaces are similar on the fine scale.
Nonetheless, IXPE’s imaging performance is robust against both mission aging and count-rate–induced effects.
We find that, after the optics realignment in June 2022, no statistically significant time slopes of the HPD are found after the optics realignment.
A marginal $\sim1$ arcsec increase of the HPD between T1 and T2 could be explained by the initial settling of the MMAs.
Linear fits over the 14 months post-realignment (June 2022 – August 2023) give slopes consistent with zero at the $<3$ sigma level 
%
After the September 2023 temperature set-point change, HPD shifted by $\leq$1 arcsec, consistent with finite thermal contraction of mirror shells and expected GPD focal-plane refocus. 
Similar stability holds versus source rate: no significant degradation at high count rates up to $\sim60$ cts s$^{-1}$. 
After the optics realignment, the average values of the HPD for Telescope 1, Telescope 2, and Telescope 3 is $26.1 \pm 0.5$ arcsec,  $32.1 \pm 0.5$ arcsec, and  $30.9 \pm 0.6$ arcsec, respectively.
After the temperature set-point change, the average value increases by $\sim1$ arcsec for each Telescope at $27.2 \pm 0.7$,  $33.3 \pm 0.7$, and $31.4 \pm 0.7$ arcsec.
IXPE is expected to fly at least until 2030, given the current trends, the projected values of the HPD by then, given current trends, will be $29.2 \pm 3.4$ arcsec, $36.2 \pm 3.6$ arcsec, and $31.0 \pm 3.5$ arcsec for the three telescopes.
%
Thus, in-flight PSF stability is confirmed: neither realignment nor temperature set-point changes have produced significant secular widening of the core or broadening of the wings. 
We also studied possible deviations of the spatially-resolved polarimetric response of the detectors.
One of these is the possible scattering-induced polarization due to mirror microscale roughness breaks that symmetry in the scattering halo, as Wolter-I geometry predicts cancellation of azimuthal Fresnel-induced polarization. 
Fresnel calculations at 2 keV ($\alpha=0.49°$) imply a maximum induced polarization $\sim10^{-4}$ that roughly halves at 7 keV.
Empirically, using ObsID 1002801 (GS 1826–238, nominally unpolarized), we measure a 99\% C.L. upper limit of 4.4\% polarization in a 2-4.5 arcmin annulus (2-6 keV). 
Therefore, from this measurement we can say that $\geq$95.6\% of grazing reflection events follow Fresnel laws.
This means that any optics-induced polarization within the PSF wings is lower than 4.4\%, and in the core is $\ll0.1\%$ (due to azimuthal cancellation). 
A second effect comes from Monte Carlo studies \citep{2013Soffitta} that predicted a lower modulation factor in the PSF wings from mis-reconstructed photoelectron tracks.
On orbit, selecting Cyg X-3 (bright, highly polarized) in 5$–$7 keV—we measure PD$_{core}$=$21.4\pm1.3\%$ versus PD$_{wings}$=$22.3\pm2.1\%$. 
The polarization, hence the modulation factor, of the wings matches that of the core within uncertainties. 
The broader, energy-dependent on-sky PSF (from mirror resolution and parallax broadening) masks the narrow-beam simulation effect because the real telescope wings contains only a few mis-reconstructed tracks, being mostly due to optics effect and not to errors in the determination of the impact points in the analysis.
Consequently, GPD reconstruction errors do not significantly degrade polarization sensitivity outside the core. 
In conclusion, these results not only validate the fidelity of IXPE’s spatially resolved polarimetry for point like and extended sources, but also affirm the telescopes’ readiness to deliver high‑precision X-ray polarization measurements throughout the mission’s planned lifetime into 2030.

\begin{acknowledgments}
The Imaging X-ray Polarimetry Explorer (IXPE) is a joint US and Italian mission. 
The US contribution is supported by the National Aeronautics and Space Administration (NASA) and led and managed by its Marshall Space Flight Center (MSFC), with industry partner Ball Aerospace (contract NNM15AA18C). 
The Italian contribution is supported by the Italian Space Agency (Agenzia Spaziale Italiana, ASI) through contract ASI-OHBI-2017-12-I.0, agreements ASI-INAF-2017-12-H0 and ASI-INFN-2017.13-H0, and its Space Science Data Center (SSDC) and by the Istituto Nazionale di Astrofisica (INAF) and the Istituto Nazionale di Fisica Nucleare (INFN) in Italy. 
This research used data products provided by the IXPE Team (MSFC, SSDC, INAF, and INFN) and distributed with additional software tools by the High-Energy Astrophysics Science Archive Research Center (HEASARC) at NASA Goddard Space Flight Center (GSFC). We kindly thank the referee for their comments and suggestions.
\end{acknowledgments}

%
\facilities{IXPE}

\software{ixpeobssim\cite{2022Baldini},
          astropy\cite{Astropy_1,Astropy_2,Astropy_3},
          leakagelib\cite{2024Dinsmore}
          }

\appendix
\label{sec:appendix}

\section{List of observations and fit results}

\begin{table}[htbp]
\digitalasset
\centering
\caption{List of the observations used in this work and results for the fits of Telescope 1 in the 2$-$3 keV energy band.
\label{tab:Telescope_1_fit_results}}
\begin{tabular}{cccccccccc}
\hline
\textbf{obsid} & \textbf{Rate} & \textbf{MET} & \textbf{Date} & \textbf{$\rm \sigma$} & \textbf{$\rm r_c$} & \textbf{$\rm \eta$} & \textbf{$r_0$} & \textbf{$\chi^2$/dof} & \textbf{HPD} \\
               & \textbf{(cts s$^{-1}$)} & \textbf{(s)} &  & \textbf{(arcsec)} & \textbf{(arcsec)} &   & \textbf{(arcsec)} &  & \textbf{(arcsec)} \\
\hline
1001899 & 0.32  & 161878860 & 2022-02-21 & 10.13 $\pm$ 0.12 & 13.13 $\pm$ 0.41 & 2.08 $\pm$ 0.06 & 159.8 $\pm$ 17.2  & 26.6/26   & 24.93 $\pm$ 0.08 \\
1004501 & 0.28  & 163479591 & 2022-03-09 & 8.31 $\pm$ 0.11  & 9.13 $\pm$ 0.51  & 1.51 $\pm$ 0.07 & 151.7 $\pm$ 26.9  & 55.5/48   & 24.89 $\pm$ 0.15 \\
1002701 & 0.30  & 164858551 & 2022-03-25 & 8.47 $\pm$ 0.11  & 9.84 $\pm$ 0.51  & 1.63 $\pm$ 0.08 & 120.4 $\pm$ 13.9  & 97.7/93   & 24.7 $\pm$ 0.11  \\
1004601 & 0.43  & 165131984 & 2022-03-28 & 8.85 $\pm$ 0.07  & 9.36 $\pm$ 0.42  & 1.59 $\pm$ 0.06 & 265.4 $\pm$ 91.4  & 42.2/43   & 23.66 $\pm$ 0.11 \\
1002801 & 6.89  & 165311011 & 2022-03-30 & 11.23 $\pm$ 0.16 & 11.2 $\pm$ 0.1   & 2.02 $\pm$ 0.02 & 109.3 $\pm$ 4.9   & 125.4/126 & 23.32 $\pm$ 0.03 \\
... & ... & ... & ... & ... & ... & ...  & ...    & ... & ...  \\
\hline 
\end{tabular}
\tablecomments{Table \ref{tab:Telescope_1_fit_results} is published in its entirety in the machine-readable format. 
A portion is shown here for guidance regarding its form and content.}
\end{table}

\begin{table}[htbp]
\digitalasset
\centering
\caption{List of the observations used in this work and results for the fits of Telescope 2 in the 2$-$3 keV energy band.
\label{tab:Telescope_2_fit_results}}
\begin{tabular}{cccccccccc}
\hline
\textbf{obsid} & \textbf{Rate} & \textbf{MET} & \textbf{Date} & \textbf{$\rm \sigma$} & \textbf{$\rm r_c$} & \textbf{$\rm \eta$} & \textbf{$r_0$} & \textbf{$\chi^2$/dof} & \textbf{HPD} \\
               & \textbf{(cts s$^{-1}$)} & \textbf{(s)} &  & \textbf{(arcsec)} & \textbf{(arcsec)} &   & \textbf{(arcsec)} &  & \textbf{(arcsec)} \\
\hline
1001899 & 0.29  & 161878860 & 2022-02-21 & 11.69 $\pm$ 0.4  & 20.91 $\pm$ 0.64 & 2.76 $\pm$ 0.1  & 132.9 $\pm$ 10.7  & 36.4/28   & 29.48 $\pm$ 0.1  \\
1004501 & 0.27  & 163479591 & 2022-03-09 & 10.2 $\pm$ 0.17  & 13.92 $\pm$ 0.91 & 1.63 $\pm$ 0.1  & 158.1 $\pm$ 25.2  & 29.5/29   & 31.18 $\pm$ 0.17 \\
1002701 & 0.29  & 164858551 & 2022-03-25 & 10.07 $\pm$ 0.13 & 13.48 $\pm$ 0.6  & 1.6 $\pm$ 0.06  & 236.8 $\pm$ 45.6  & 21.6/24   & 30.47 $\pm$ 0.12 \\
1004601 & 0.43  & 165131984 & 2022-03-28 & 10.34 $\pm$ 0.18 & 15 $\pm$ 0.71    & 1.89 $\pm$ 0.08 & 301.3 $\pm$ 71.9  & 21/24     & 29.08 $\pm$ 0.13 \\
1002801 & 6.46  & 165311011 & 2022-03-30 & 14.7 $\pm$ 0.43  & 20.84 $\pm$ 0.37 & 2.88 $\pm$ 0.06 & 93.7 $\pm$ 3.7    & 174.2/147 & 29.63 $\pm$ 0.04 \\
... & ... & ... & ... & ... & ... & ...  & ...    & ... & ...  \\
\hline 
\end{tabular}
\tablecomments{Table \ref{tab:Telescope_2_fit_results} is published in its entirety in the machine-readable format. 
A portion is shown here for guidance regarding its form and content.}
\end{table}
\begin{table}[htbp]
\digitalasset
\centering
\caption{List of the observations used in this work and results for the fits of Telescope 3 in the 2$-$3 keV energy band.
\label{tab:Telescope_3_fit_results}}
\begin{tabular}{cccccccccc}
        \hline
        \textbf{obsid} & \textbf{Rate} & \textbf{MET} & \textbf{Date} & \textbf{$\rm \sigma$} & \textbf{$\rm r_c$} & \textbf{$\rm \eta$} & \textbf{$r_0$} & \textbf{$\chi^2$/dof} & \textbf{HPD} \\
                       & \textbf{(cts s$^{-1}$)} & \textbf{(s)} &  & \textbf{(arcsec)} & \textbf{(arcsec)} &   & \textbf{(arcsec)} &  & \textbf{(arcsec)} \\
        \hline
1001899 & 0.41  & 161878860 & 2022-02-21 & 11.67 $\pm$ 0.11 & 20.63 $\pm$ 0.64 & 2.43 $\pm$ 0.08  & 185.7 $\pm$ 20     & 34.6/31   & 30.73 $\pm$ 0.08  \\
1004501 & 0.25  & 163479591 & 2022-03-09 & 9.83 $\pm$ 0.2   & 15.6 $\pm$ 0.9   & 1.95 $\pm$ 0.11  & 279.1 $\pm$ 85.9   & 49.1/43   & 28.19 $\pm$ 0.15  \\
1002701 & 0.27  & 164858551 & 2022-03-25 & 10.07 $\pm$ 0.2  & 10.26 $\pm$ 1.08 & 2.3 $\pm$ 0.05   & 984.4 $\pm$ 1537.2 & 18/18     & 29.54 $\pm$ 0.13  \\
1004601 & 0.39  & 165131984 & 2022-03-28 & 9.67 $\pm$ 0.15  & 14.21 $\pm$ 0.69 & 1.84 $\pm$ 0.09  & 226.7 $\pm$ 56.4   & 57.5/43   & 27.61 $\pm$ 0.14  \\
1002801 & 6.16  & 165311011 & 2022-03-30 & 11.22 $\pm$ 0.19 & 18.78 $\pm$ 0.35 & 2.57 $\pm$ 0.06  & 105.2 $\pm$ 5.1    & 121.5/118 & 28.4 $\pm$ 0.03   \\
... & ... & ... & ... & ... & ... & ...  & ...    & ... & ...  \\
\hline 
\end{tabular}
\tablecomments{Table \ref{tab:Telescope_3_fit_results} is published in its entirety in the machine-readable format. 
A portion is shown here for guidance regarding its form and content.}
\end{table}

\bibliography{Arxiv_version}{}
\bibliographystyle{aasjournalv7}



\end{document}